\documentclass[twocolumn,aps,prb,floats,showpacs]{revtex4}

\usepackage{bm}
\usepackage{dcolumn}
\usepackage{epsfig}

\begin{document}

\title{Universal parametric correlations in the classical limit of quantum transport}

\author{Piet W.\ Brouwer$^{1,2}$ and Saar Rahav$^3$}

\affiliation{$^1$ Laboratory of Atomic and Solid State Physics, Cornell
  University, Ithaca, NY 14853, USA \\
  $^2$ Arnold Sommerfeld Center for Theoretical Physics, Ludwig-Maximilians-Universit\"at, 80333 M\"unchen, Germany \\
$^3$ Department of Chemistry and Biochemistry, University of Maryland,
College Park, MD 20742, USA}

\date{\today}


\begin{abstract}
Quantum corrections to transport through a chaotic ballistic cavity are known to be universal. The universality not only applies to the magnitude of quantum corrections, but also to their dependence on external parameters, such as the Fermi energy or an applied magnetic field. Here we consider such parameter dependence of quantum transport in a ballistic chaotic cavity in the semiclassical limit obtained by sending $\hbar \to 0$ without changing the classical dynamics of the open cavity. In this limit quantum corrections are shown to have a universal parametric dependence which is not described by random matrix theory.
\end{abstract}

\maketitle

Central to the field of `quantum chaos' is the observation that statistical fluctuations of the spectra of quantum systems whose classical dynamics is chaotic are universal, as well as the relation between the universal spectral fluctuations and random matrix theory.\cite{kn:haake1991} The universality not only applies to probability distributions of energy levels, but also includes correlations at different values of external parameters, such as an applied magnetic field.\cite{kn:altshuler1995,kn:guhr1998} A necessary condition for the existence of universal spectral statistics is that the time $\tau_{\rm erg}$ needed for ergodic exploration of the phase space be much smaller than the Heisenberg time $\tau_{\rm H} = 2 \pi \hbar/\Delta$, $\Delta$ being the mean spacing between energy levels. Since $\tau_{\rm erg}$ is a classical time scale, whereas $\tau_{\rm H}$ involves Planck's constant $\hbar$, the condition $\tau_{\rm erg} \ll \tau_{\rm H}$ is equivalent to the semiclassical limit $\hbar \to 0$.\cite{foot1}

Similar considerations apply to open quantum systems,\cite{kn:beenakker1997} for which the role of energy levels is played by the transport coefficients (or by the scattering matrix). A prototypical example of an open quantum system with chaotic classical dynamics is an electron in a two-dimensional ballistic cavity coupled to electron reservoirs via ballistic contacts.\cite{kn:bluemel1988,kn:jalabert1990} Such cavities, or `quantum dots', can be realized experimentally in semiconductor heterostructures.\cite{kn:kouwenhoven1997} In this context, `universality' means that the statistical fluctuations of the transport coefficients do not depend on the shape of the cavity, as long as the classical dynamics is chaotic. 

In addition to $\tau_{\rm erg}$ and $\tau_{\rm H}$, an open cavity has a third characteristic time scale, the mean dwell time $\tau_{\rm D}$. The appearance of a third time scale complicates the conditions for the applicability of random matrix theory (RMT), as well as the relation to the semiclassical limit $\hbar \to 0$. The reason is that the condition necessary for universal quantum transport,\cite{foot2}
\begin{eqnarray}
  \tau_{\rm erg} \ll \tau_{\rm D} \ll \tau_{\rm H}, \label{eq:condopen}
\end{eqnarray}
is not sufficient for the applicability of RMT.\cite{kn:aleiner1996} The condition for RMT involves the Ehrenfest time $\tau_{\rm E}$, which for a two-dimensional cavity reads\cite{kn:larkin1968,kn:zaslavsky1981}
\begin{equation}
  \tau_{\rm E} = \lambda^{-1} \ln(\tau_{\rm H}/\tau_{\rm erg}),
\end{equation}
where $\lambda \sim \tau_{\rm erg}^{-1}$ is the Lyapunov exponent of the cavity's classical dynamics. The Ehrenfest time is the minimal dwell time necessary for quantum interference,\cite{kn:aleiner1996} hence RMT applies only if $\tau_{\rm E} \ll \tau_{\rm D}$, {\em i.e.}, if
\begin{equation}
  \tau_{\rm erg} \ln (\tau_{\rm H}/\tau_{\rm erg}) \ll \tau_{\rm D}.
  \label{eq:cond}
\end{equation}

The condition (\ref{eq:cond}) has little impact on most experiments on ballistic quantum dots, for which the logarithm $\ln(\tau_{\rm H}/\tau_{\rm erg})$ is not numerically large.\cite{kn:kouwenhoven1997} Nevertheless, since $\tau_{\rm H}/\tau_{\rm erg}$ is proportional to $\hbar^{-1}$, it has important consequences for the relation between RMT and the semiclassical limit $\hbar \to 0$ in an open cavity. Obeying the condition (\ref{eq:cond}) while sending $\hbar \to 0$ is possible only if the ratio $\tau_{\rm D}/\tau_{\rm erg}$ grows at least logarithmically with $\hbar$. Since both $\tau_{\rm erg}$ and $\tau_{\rm D}$ are classical time scales, this means that RMT describes the cavity's transport coefficients in the limit $\hbar \to 0$ only if the classical dynamics of the open cavity is modified in the limiting process. 

The last decade has shown an increased interest in the opposite limit, obtained by sending $\hbar \to 0$ at fixed $\tau_{\rm erg}$ and $\tau_{\rm D}$ before taking the limit $\tau_{\rm erg}/\tau_{\rm D} \to 0$.\cite{kn:aleiner1996,kn:agam2000,kn:adagideli2003,kn:vavilov2003,kn:tworzydlo2004,kn:schomerus2004} In this case the Ehrenfest time $\tau_{\rm E} \gg \tau_{\rm D}$. We refer to this limit as the `true semiclassical limit', because it involves sending $\hbar \to 0$ without changing the classical dynamics of the open cavity. Although some quantum effects cease to exist in the true semiclassical limit (examples are the shot noise power\cite{kn:beenakker1991c,kn:agam2000} and the ensemble average $\langle \delta G \rangle$ of the quantum correction $\delta G$ to the cavity's conductance\cite{kn:aleiner1996,kn:adagideli2003}), not all quantum effects disappear. This remarkable observation was first made for the conductance fluctuations, whose mean square $\langle \delta G^2 \rangle$ remains equal to the RMT prediction in the true semiclassical limit.\cite{kn:tworzydlo2004,kn:brouwer2006} In this communication, we consider correlations between conductances at different external parameters, such as the Fermi energy or an applied magnetic field. In the true semiclassical limit we find a result that is universal, but with a functional dependence on external parameters that differs from random matrix theory.

The parametric dependence of the conductance fluctuations is described by the conductance autocorrelation function $\langle \delta G(\varepsilon,b) \delta G(\varepsilon',b') \rangle$, where $\varepsilon$ and $b$ are properly normalized energy and magnetic field.\cite{kn:beenakker1997} Our calculation, which is outlined below, gives
\begin{equation}
  \langle \delta G(\varepsilon,b) \delta G(\varepsilon',b') \rangle =
  P_1^2 P_2^2 \sum_{\pm} 
  \mbox{Re}\, {\cal D}_{\pm},
  \label{eq:dGclass}
\end{equation}
where $\delta G$ is measured in units of $2e^2/h$,
$P_1$ and $P_2$ are the classical probabilities that an electron in the cavity escapes through contacts $1$ or $2$, respectively, and
\begin{equation}
  {\cal D}^{-1}_{\pm} = 1 - i (\varepsilon-\varepsilon') + (1/2)
  (b \pm b')^2.
  \label{eq:Ddef}
\end{equation}
The RMT prediction has $\mbox{Re}\, {\cal D}_{\pm}$ replaced by $|{\cal D}_{\pm}|^2$. In particular, at $\varepsilon=\varepsilon'$ and $|b|$, $|b'|\gg 1$, the conductance autocorrelation in the true semiclassical limit has a Lorentzian dependence on the magnetic field difference $b-b'$, whereas RMT predicts a Lorentzian squared.\cite{kn:jalabert1990,kn:efetov1995,kn:frahm1995a} Generalization of our result to other parameters $x_j$ that, {\em e.g.}, represent a small deformation of the cavity's shape amounts to the replacement of Eq.\ (\ref{eq:Ddef}) by
\begin{equation}
  {\cal D}_{\pm}^{-1} = 1 - i (\varepsilon-\varepsilon') 
  + \frac{1}{2} (b\pm b')^2 + \frac{1}{2} \sum_{j} (x_j - x_j')^2,
  \label{eq:Ddeflong}
\end{equation}
where it is assumed that the parameters $x_j$ do not break time-reversal symmetry.

The derivation of Eq.\ (\ref{eq:dGclass}) closely follows the calculation of the variance of the conductance of a ballistic cavity, which is described in Ref.\ \onlinecite{kn:brouwer2006}. That calculation starts from the relation between the conductance autocorrelation function and the cavity's reflection coefficients $R_1$ and $R_2$,
\begin{equation}
  \langle \delta G(\varepsilon,b) \delta G(\varepsilon',b') \rangle =
  \langle \delta R_1(\varepsilon,b) \delta R_2(\varepsilon',b')\rangle,
  \label{eq:GR}
\end{equation}
together with an expression that relates $R_j$ to a double sum over classical trajectories $\alpha_j$ and $\beta_j$ that begin and end at contact $j$, $j=1,2$,\cite{kn:jalabert1990}
\begin{equation}
  R_j = 
  \sum_{\alpha_j,\beta_j} A_{\alpha_j} A_{\beta_j}e^{i({\cal S}_{\alpha_j} - {\cal S}_{\beta_j})/\hbar}
  ,\ \ j=1,2.
  \label{eq:rclass}
\end{equation}
Here $A$ and ${\cal S}$ are the stability amplitude and classical action of the trajectories. Phase shifts from reflections off the cavity boundary are absorbed into the definition of the action. Upon entry and exit, the two classical trajectories $\alpha_j$ and $\beta_j$ have transverse momenta $|p_{\perp,\alpha_j}|=|p_{\perp,\beta_j}|$ compatible with the quantized modes in the contacts.\cite{kn:jalabert1990}

Upon using Eqs.\ (\ref{eq:GR}) and (\ref{eq:rclass}), the conductance autocorrelation function is expressed as a quadruple sum over classical trajectories $\alpha_1$, $\beta_1$, $\alpha_2$, and $\beta_2$. Only combinations of four trajectories for which the total action difference ${\cal S}_{\alpha_1} - {\cal S}_{\beta_1} + {\cal S}_{\alpha_2} - {\cal S}_{\beta_2}$ is of order $\hbar$ systematically contribute to the autocorrelation function. Such small action differences occur only if the trajectories $\alpha_1$ and $\alpha_2$, on the one hand, and the trajectories $\beta_1$ and $\beta_2$, on the other hand, are piecewise identical, up to classical phase space distances of order $\hbar^{1/2}$ or less.\cite{kn:aleiner1996,kn:richter2002,kn:heusler2006}

\begin{figure}
\epsfxsize=0.95\hsize
\epsffile{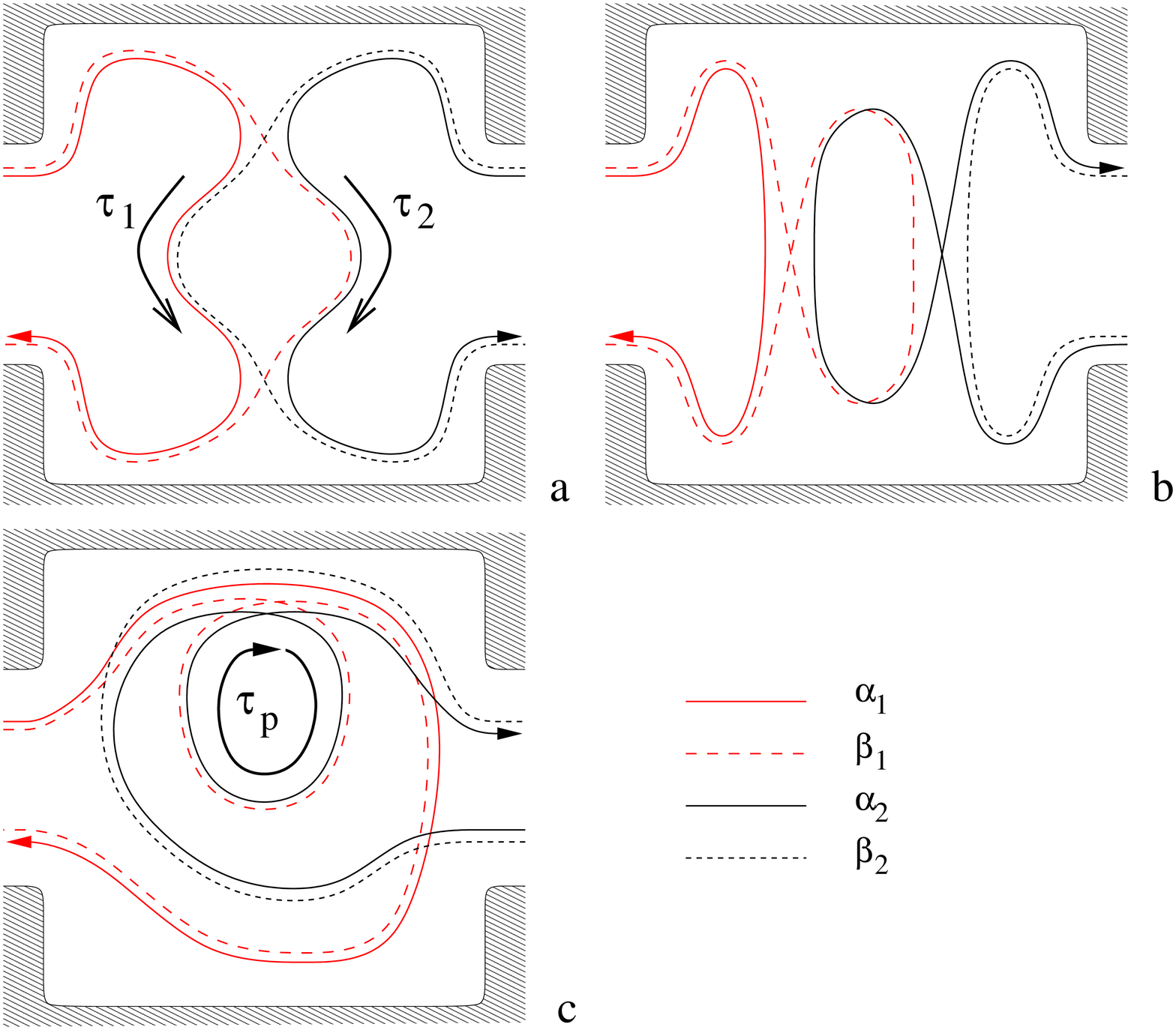}
\caption{\label{fig:2} (Color online) Schematic drawing of quadruples of trajectories that contribute to the conductance autocorrelation function. The true trajectories are piecewise straight, with specular reflection at the boundaries.}
\end{figure}

There are two general classes of trajectories that meet these criteria. They are shown schematically in Figs.\ \ref{fig:2}a and b. Both classes of trajectories have their counterpart in the diagrammatic theory of conductance fluctuations in disordered metals.\cite{kn:altshuler1986} In Fig.\ \ref{fig:2}a, the four trajectories have two separate small-angle encounters. Outside the encounters, the trajectories $\alpha_1$ and $\beta_1$, and $\alpha_2$ and $\beta_2$ are paired. Between the encounters $\alpha_1$ is paired with $\beta_2$ and $\alpha_2$ is paired with $\beta_1$. The duration of the encounters is long enough that the total action difference is of order $\hbar$. This is achieved if the encounter duration, defined as the time that the phase space distance between the four trajectories is less than a certain classical cut-off, is the Ehrenfest time $\tau_{\rm E}$ or longer.\cite{kn:aleiner1996,kn:richter2002,kn:heusler2006} In Fig.\ \ref{fig:2}b, the trajectories $\alpha_1$ and $\beta_1$ are identical up to a closed loop, which is in $\beta_1$ but not in $\alpha_1$. The same closed loop is also the difference of $\beta_2$ and $\alpha_2$. There is a second possibility, different from the first one by complex conjugation, in which the closed loop is part of $\alpha_1$ and $\beta_2$, but not $\alpha_2$ and $\beta_1$. Only the first possibility is shown in the figure. Although the four trajectories shown in Fig.\ \ref{fig:2}b represent the generic case of interfering trajectories where the two trajectories in each pair differ by a closed loop, for a chaotic cavity such quadruplets contribute to $\langle \delta G^2 \rangle$ only if $\alpha_1$, $\beta_1$, $\alpha_2$, and $\beta_2$ meet the closed loop in a single small-angle encounter of all four trajectories,\cite{kn:brouwer2006} see Fig.\ \ref{fig:2}c. 
In the presence of time-reversal symmetry, two additional contributions to the conductance autocorrelation function appear, which are obtained by time-reversing the trajectories $\alpha_2$ and $\beta_2$ in Figs.\ \ref{fig:2}a--c.
 
The external parameters enter the calculation of the conductance autocorrelation function through the parameter dependence of the classical actions.\cite{kn:jalabert1990} Since the actions of the trajectory pairs ($\alpha_1$, $\beta_1$) and ($\alpha_2$,$\beta_2$) are taken at equal values of the parameters, all parametric dependence must arise from action differences accumulated when the trajectories in these pairs are separated. For the trajectories in Fig.\ \ref{fig:2}a this occurs during the two stretches of duration $\tau_1$ and $\tau_2$ between the encounters; For the trajectories in Fig.\ \ref{fig:2}c this is during the closed loop, the period of which is denoted $\tau_{\rm p}$. Since we sum over all trajectories, it is sufficient to know the mean and variance of these action differences,
\begin{eqnarray}
  \langle {\cal S}(\varepsilon,b) - {\cal S}(\varepsilon',b') \rangle &=& 
  \hbar (\varepsilon-\varepsilon') \tau/\tau_{\rm D},\nonumber \\
  \langle ({\cal S}(\varepsilon,b) - {\cal S}(\varepsilon',b'))^2
  \rangle &=& \hbar^2 (b-b')^2\tau/\tau_{\rm D}, 
  \label{eq:Smean}
\end{eqnarray}
where $\tau=\tau_1$, $\tau_2$, or $\tau_{\rm p}$ is the duration of the stretch of the trajectories over which the action difference is accumulated. The action difference between a trajectory and its time-reversed is obtained by replacing $b'$ by $-b'$. The rescaled energy $\varepsilon$ is the energy measured in units of $\hbar/\tau_{\rm D}$. Equation (\ref{eq:Smean}) should be seen as the definition of the rescaled fields $b$ and $b'$; up to a numerical constant that depends on the cavity shape one has $b \sim (e \Phi/\hbar c) (\tau_{\rm D}/\tau_{\rm erg})^{1/2}$, $\Phi$ being the magnetic flux through the cavity.\cite{kn:jalabert1990}

The sum over all classical trajectories, but without parametric dependence, has been calculated before.\cite{kn:brouwer2006} In order to obtain the full parametric dependence we take the trajectory sum before the final integration over the times $\tau_1$, $\tau_2$, and $\tau_p$ from Ref.\ \onlinecite{kn:brouwer2006},
\begin{eqnarray}
  \langle \delta G^2 \rangle &=& 2 
  e^{- 2 \tau_{\rm E}/\tau_{\rm D}}
  P_1^2 P_2^2 \int_0^{\infty} \frac{d\tau_1 d\tau_2}{\tau_{\rm D}^2} 
  e^{-(\tau_1+\tau_2)/\tau_{\rm D}} \nonumber
  \\
  && \mbox{} +
  2 (1-e^{- 2 \tau_{\rm E}/\tau_{\rm D}}) P_1^2 P_2^2 
  \int_0^{\infty} \frac{d\tau_{\rm p}}{\tau_{\rm D}} 
  e^{-\tau_{\rm p}/\tau_{\rm D}}. \nonumber \\
  \label{eq:dGt}
\end{eqnarray}
The first term, with the double integration over $\tau_1$ and $\tau_2$, originates from the trajectory class of Fig.\ \ref{fig:2}a; The second term, with the single integration over $\tau_{\rm p}$, comes from Fig.\ \ref{fig:2}c. Both terms include time-reversed contributions. The second term also includes the complex conjugate contribution not shown in Fig.\ \ref{fig:2}c. Taking Eq.\ (\ref{eq:dGt}) as our starting point, we find the parametric dependence of the conductance autocorrelation function upon insertion of the appropriate factors $\langle \exp(i \Delta {\cal S}/\hbar) \rangle$ for each of the stretches where action differences are accumulated. Each time integration in Eq.\ (\ref{eq:dGt}) then gives a factor ${\cal D}_-$ or ${\cal D}_-^*$, where ${\cal D}_{-}$ is defined in Eq.\ (\ref{eq:Ddef}) above. Time integrations involving time-reversed trajectories give a factor ${\cal D}_+$ or ${\cal D}_+^*$. We thus find
\begin{eqnarray}
  \langle \delta G(\varepsilon,b) \delta G(\varepsilon',b') \rangle &=&
  P_1^2 P_2^2 \sum_{\pm} \left[
  e^{-2\tau_{\rm E}/\tau_{\rm D}} |{\cal D}_{\pm}|^2
  \right. \nonumber \\ && \left. \mbox{}
  + (1 - e^{-2\tau_{\rm E}/\tau_{\rm D}})\, \mbox{Re}\, {\cal D}_{\pm}
  \right]. ~~
  \label{eq:Gresult}
\end{eqnarray}
The true semiclassical limit corresponds to the limit $\tau_{\rm E}/\tau_{\rm D} \to \infty$. In this limit, only trajectories of the type shown in Fig.\ \ref{fig:2}c contribute to the conductance autocorrelation function. RMT is recovered in the opposite limit $\tau_{\rm E}/\tau_{\rm D} \to 0$. The different parametric dependences in the two limits reflect the different number of time integrations involved in the contributions of Fig.\ \ref{fig:2}a and c. 

\begin{figure}
\epsfxsize=0.8\hsize
\epsffile{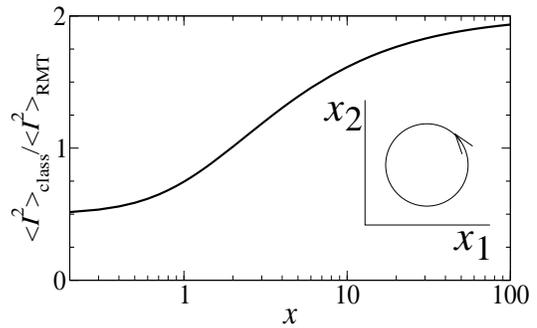}
\caption{\label{fig:3} Ratio of mean square pumped current in the true semiclassical limit $\langle I^2 \rangle_{\rm class}$ and the RMT prediction $\langle I^2 \rangle_{\rm RMT}$.
Inset: Pumping contour in the $(x_1,x_2)$ plane. The ratio shown in the main figure is calculated for a circular pumping contour with radius $x$.}
\end{figure}

Another noteworthy example of an observable that measures the universal parametric dependence of quantum transport is the current $I$ through a `quantum pump', a chaotic cavity with two parameters that are varied periodically in time.\cite{kn:brouwer1998,kn:switkes1999,kn:zhou1999} In an experimental realization, these parameters would be two gate voltages that determine the shape of a semiconductor quantum dot.\cite{kn:switkes1999}
The rescaled parameters that determine the magnitude of the pumped current are the same as those that appear in the conductance autocorrelation function. Hence, a measurement of the mean and variance of the pumped current is a direct test of the universality of quantum transport and involves no further scaling factors.

In the adiabatic limit (frequency $\omega \ll \tau_{\rm D}^{-1}$), the time-averaged current $I_j$ through contact $j$, $j=1,2$, for a cavity with time-dependent parameters $x_1$ and $x_2$ can be expressed in terms of an integral over the area ${\cal A}$ enclosed in the $(x_1,x_2)$ plane in one cycle.\cite{kn:brouwer1998}, see Fig.\ \ref{fig:3} The integrand is expressed in terms of classical trajectories connecting the two contacts to the cavity in a manner very similar to Eq.\ (\ref{eq:rclass}) above,\cite{kn:martinez-mares2004}
\begin{eqnarray}
  I_j &=& 2 e \omega \int_{\cal A} dx_1 dx_2 \Pi_j(x_1,x_2), \nonumber \\
  \Pi_j &=& 
  \sum_{\alpha,\beta} \frac{A_{\alpha} A_{\beta}}{(2 \pi \hbar)^2}
  \frac{\partial {\cal S}_{\alpha}}{\partial x_1} 
  \frac{\partial {\cal S}_{\beta}}{\partial x_2}
  \sin\left( \frac{{\cal S}_{\alpha} - {\cal S}_{\beta}}{\hbar} \right).
\end{eqnarray}
Here the trajectories $\alpha$ and $\beta$ exit through contact $j$, $j=1,2$, but they may enter the cavity through either contact. As in Eq.\ (\ref{eq:rclass}), $\alpha$ and $\beta$ have transverse momenta $|p_{\perp,\alpha}| = |p_{\perp,\beta}|$ upon entrance and exit that are compatible with the quantized modes in the contacts.
Performing the summation over classical trajectories, one finds that the ensemble average $\langle \Pi(x_1,x_2) \rangle = 0$, whereas the mesoscopic fluctuations are given by
\begin{eqnarray}
  \langle \Pi_j \Pi_j \rangle &=&
  -
  \frac{P_1 P_2}{64 \pi^4}
  \left( \frac{\partial^2}{\partial x_1^2} + \frac{\partial^2}{\partial x_2^2}
  \right)
  \left[ e^{-2 \tau_{\rm E}/\tau_{\rm D}} |{\cal D}_-|^2
  \right. \nonumber \\ && \left. \mbox{}
  + (1 - e^{-2 \tau_{\rm E}/\tau_{\rm D}})\, \mbox{Re}\, {\cal D}_-
  \right],
\end{eqnarray}
where ${\cal D}_-$ is given by Eq.\ (\ref{eq:Ddeflong}) above and the primed parameters refer to the second factor of the kernel $\Pi$.  Again, the limit $\tau_{\rm E}/\tau_{\rm D} \to 0$ agrees with the RMT prediction,\cite{kn:shutenko2000} whereas the true semiclassical limit $\tau_{\rm E}/\tau_{\rm D} \to \infty$ gives different, but still universal parametric correlations for the pumped current. For small pumping amplitudes (variation of the dimensionless parameters $x_{1,2}$ much less than unity), $\langle I^2 \rangle$ in the true semiclassical limit is half the RMT prediction.\cite{kn:rahav2006c} However, for large amplitudes, the pumped current in the semiclassical limit is larger than the RMT prediction. This is illustrated in Fig.\ \ref{fig:3}, where we have shown the ratio of the mean square current in the true semiclassical limit and the RMT prediction for a harmonic time dependence of the parameters, $x_1(t) = x \sin(\omega t)$ and $x_2(t) = x \cos(\omega t)$.

In conclusion, we considered the parameter dependence of the conductance and the pumped current in an open chaotic cavity in the `true semiclassical limit', defined as the limit $\hbar \to 0$ at fixed classical dynamics of the open cavity. Although it was known that certain quantum interference corrections survive in this limit,\cite{kn:tworzydlo2004,kn:brouwer2006,kn:rahav2006c} the parametric correlations considered here manifestly show that quantum transport in the true semiclassical limit is universal, but not described by random matrix theory. Thus, the true semiclassical limit is identified as a nontrivial regime of universal quantum transport, separate from random matrix theory. In this respect, open ballistic cavities are different from closed cavities, for which spectral statistics always agree with random matrix theory in the limit $\hbar \to 0$. 

We are grateful to I.~Aleiner, A.~Altland, and U.~Smilansky for useful discussions. This work was supported by the Packard Foundation, the Humboldt Foundation, and by the NSF under grant no.\ 0334499. We gratefully acknowledge the hospitality of the Lewiner Institute at the Technion, Israel, where a significant part of this work was performed.


\end{document}